\documentclass[aps,showpacs,twocolumn,floats,superscriptaddress,floatfix]{revtex4}
\usepackage{graphicx,bm,times,url}
\usepackage{amsmath}
\graphicspath{{./fig/}{./png/}}
\newcommand{\BoldVec}[1]{\mathchoice%
  {\mbox{\boldmath $\displaystyle     #1$}}%
  {\mbox{\boldmath $\textstyle        #1$}}%
  {\mbox{\boldmath $\scriptstyle      #1$}}%
  {\mbox{\boldmath $\scriptscriptstyle#1$}}%
}
\newcommand{\EQ}{\begin{equation}}
\newcommand{\EN}{\end{equation}}
\newcommand{\EQA}{\begin{eqnarray}}
\newcommand{\ENA}{\end{eqnarray}}
\newcommand{\eq}[1]{(\ref{#1})}

\newcommand{\Eq}[1]{Eq.~(\ref{#1})}

\newcommand{\Sec}[1]{\S\,\ref{#1}}

\newcommand{\Fig}[1]{Fig.~\ref{#1}}

\newcommand{\bra}[1]{\langle #1\rangle}

\newcommand{\UU}{\BoldVec{U} {}}

\newcommand{\BB}{\BoldVec{B} {}}

\newcommand{\AAA}{\BoldVec{A} {}}

\newcommand{\JJ}{\BoldVec{J} {}}

\newcommand{\FF}{\BoldVec{F} {}}

\newcommand{\SSS}{\BoldVec{S} {}}

\newcommand{\nab}{\BoldVec{\nabla} {}}

\newcommand{\SSSS}{\bm{\mathsf{S}}}

\newcommand{\DD}{{\rm D} {}}
\newcommand{\dd}{{\rm d} {}}

\def\Pm{\mbox{\rm Pr}_{\rm M}}

\def\cs{c_{\rm s}}

\def\half{{\textstyle{1\over2}}}

\def\onethird{{\textstyle{1\over3}}}

\newcommand{\yan}[3]{, Astron. Nachr. {\bf #2}, #3 (#1).}

\newcommand{\ysph}[3]{, Sol.\ Phys. {\bf #2}, #3 (#1).}

\newcommand{\yjfm}[3]{, J. Fluid Mech. {\bf #2}, #3 (#1).}

\newcommand{\yprd}[3]{, Phys.\ Rev.\ D {\bf #2}, #3 (#1).}

\newcommand{\yprl}[3]{, Phys.\ Rev.\ Lett.\ {\bf #2}, #3 (#1).}

\newcommand{\yjgr}[3]{, J. Geophys. Res. {\bf #2}, #3 (#1).}

\newcommand{\yapj}[3]{, Astrophys. J. {\bf #2}, #3 (#1).}

\newcommand{\yapjl}[3]{, Astrophys. J. {\bf #2}, #3 (#1).}
\newcommand{\ypp}[3]{, Phys. Plasmas {\bf #2}, #3 (#1).}

\newcommand{\ypf}[3]{, Phys. Fluids {\bf #2}, #3 (#1).}

\newcommand{\ygafd}[3]{, Geophys. Astrophys. Fluid Dyn. {\bf #2}, #3 (#1).}

\newcommand{\yjour}[4]{, #2 {\bf #3}, #4 (#1).}

\begin{document}
\preprint{NORDITA 2009-66}

\title{Magnetic field decay of three interlocked flux rings with zero linking number}

\author{Fabio Del Sordo}

\author{Simon Candelaresi}

\author{Axel Brandenburg}

\affiliation{NORDITA, AlbaNova University Center, Roslagstullsbacken 23, SE-10691 Stockholm, Sweden}
\affiliation{Department of Astronomy,
Stockholm University, SE 10691 Stockholm, Sweden}

\date{Received 22 October 2009; revised 27 January 2010; published 3 March 2010}
\begin{abstract}
The resistive decay of chains of three interlocked magnetic flux
rings is considered.
Depending on the relative orientation of the magnetic field in the three
rings, the late-time decay can be either fast or slow.
Thus, the qualitative degree of tangledness is less important
than the actual value of the linking number or, equivalently,
the net magnetic helicity.
Our results do not suggest that invariants of higher order than that of
the magnetic helicity need to be considered to characterize the decay
of the field.
\end{abstract}

\maketitle

\section{Introduction}
\label{Introduction}

Magnetic helicity plays an important role in
plasma physics \cite{Tay74,JC84,BF84},
solar physics \cite{Low96,RK94,RK96},
cosmology \cite{BEO96,FC00,CHB05},
and dynamo theory \cite{PFL76,BS05}.
This is connected with the fact that magnetic helicity is a conserved
quantity in ideal magnetohydrodynamics  \cite{Wol58}.
The conservation law of magnetic helicity
is ultimately responsible for inverse cascade behavior
that can be relevant for spreading primordial magnetic field over
large length scales.
It is also likely the reason why the magnetic fields of many
astrophysical bodies have length scales that are larger than
those of the turbulent motions responsible for driving these fields.
In the presence of finite magnetic diffusivity, the magnetic helicity
can only change on a resistive time scale.
Of course, astrophysical bodies are open, so magnetic helicity can change
by magnetic helicity fluxes out of or into the domain of interest.
However, such cases will not be considered in the present paper.

In a closed or periodic domain without external energy supply,
the decay of a magnetic field depends
critically on the value of the magnetic helicity.
This is best seen by considering spectra of magnetic energy
and magnetic helicity.
The magnetic energy spectrum $M(k)$ is normalized such that
\EQ
\int M(k)\,d k=\bra{\BB^2}/2\mu_0,
\EN
where $\BB$ is the magnetic field, $\mu_0$ is the magnetic permeability,
and $k$ is the wave number (ranging from 0 to $\infty$).
The magnetic helicity spectrum $H(k)$ is normalized such that
\EQ
\int H(k)\,d k=\bra{\AAA\cdot\BB},
\EN
where $\AAA$ is the magnetic vector potential with $\BB=\nab\times\AAA$.
In a closed or periodic domain, $H(k)$ is gauge-invariant, i.e.\ it does not change
after adding a gradient term to $\AAA$.
For finite magnetic helicity, the magnetic energy spectrum is bound
from below \cite{Wol58} such that
\EQ
M(k)\ge k|H(k)|/2\mu_0.
\label{bound}
\EN
This relation is also known as the realizability condition \cite{Mof69}.
Thus, the decay of a magnetic field is subject to a
corresponding decay of its associated magnetic helicity.
Given that in a closed or periodic domain the magnetic helicity
changes only on resistive time scales \cite{Ber84},
the decay of magnetic energy is slowed down correspondingly.
More detailed statements can be made about the decay of turbulent magnetic
fields, where the energy decays in a power-law fashion proportional to
$t^{-\sigma}$.
In the absence of magnetic helicity, $\bra{\AAA\cdot\BB}=0$, we have
a relatively rapid decay with $\sigma\approx1.3$ \cite{MacLow}, while with
$\bra{\AAA\cdot\BB}\neq0$, the decay is slower with $\sigma$ between
1/2 \cite{CHB05} and  2/3 \cite{BM99}.

The fact that the decay is slowed down in the helical case
is easily explained in terms
of the topological interpretation of magnetic helicity.
It is well known that the magnetic helicity can be expressed in
terms of the linking number $n$ of discrete magnetic flux
ropes via \cite{Mof69}
\EQ
\int\AAA\cdot\BB\,\dd V=2n\Phi_1\Phi_2,
\EN
where
\EQ
\Phi_i=\int_{S_i}\BB\cdot\dd\SSS\quad\mbox{(for $i=1$ and 2)}
\label{FluxDef}
\EN
are the magnetic fluxes of the two ropes
with cross-sectional areas $S_1$ and $S_2$.
The slowing down of the decay is then plausibly explained by the fact that a
decay of magnetic energy is connected with a decay of magnetic helicity via the
realizability condition \eqref{bound}.
Thus, a decay of magnetic helicity can be achieved either
by a decay of the magnetic flux or by magnetic reconnection.
Magnetic flux can decay through annihilation with oppositely oriented flux.
Reconnection on the other hand reflects a change in the topological
connectivity, as demonstrated in detail in Ref.~\cite[p.28]{PriestRecon2000}.

\begin{figure}[t!]\begin{center}
\includegraphics[width=.25\columnwidth]{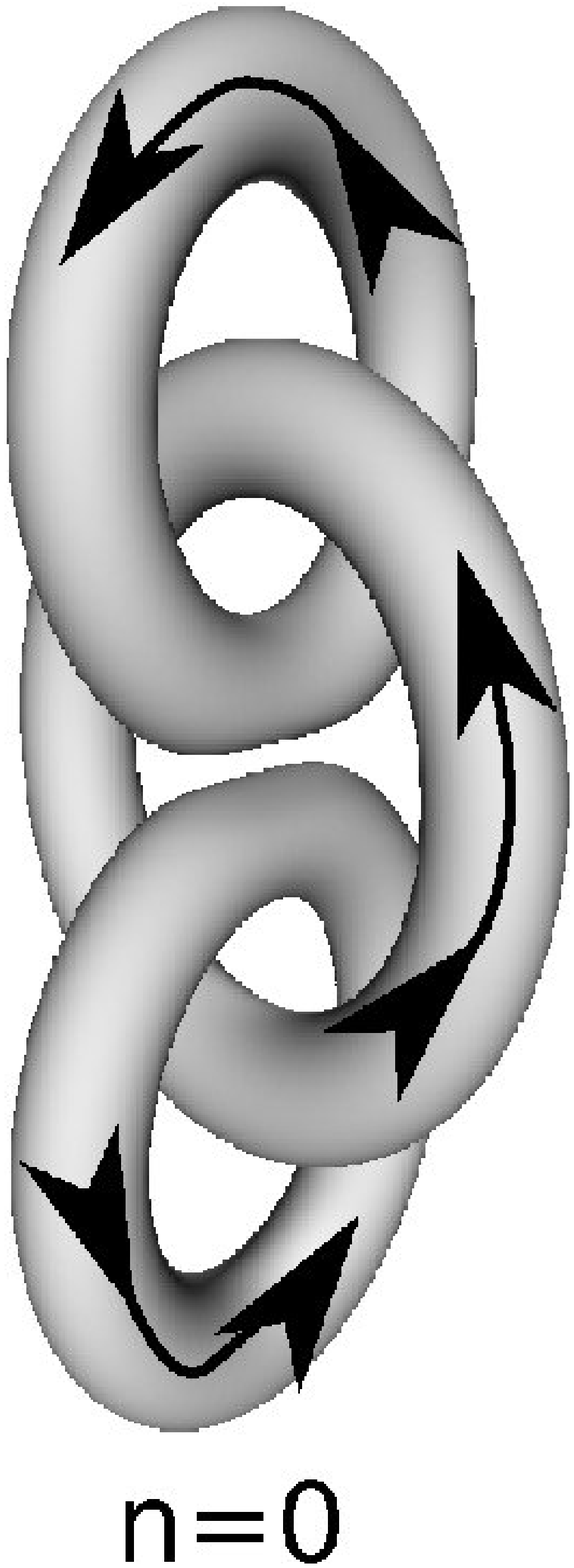} \qquad
\includegraphics[width=.25\columnwidth]{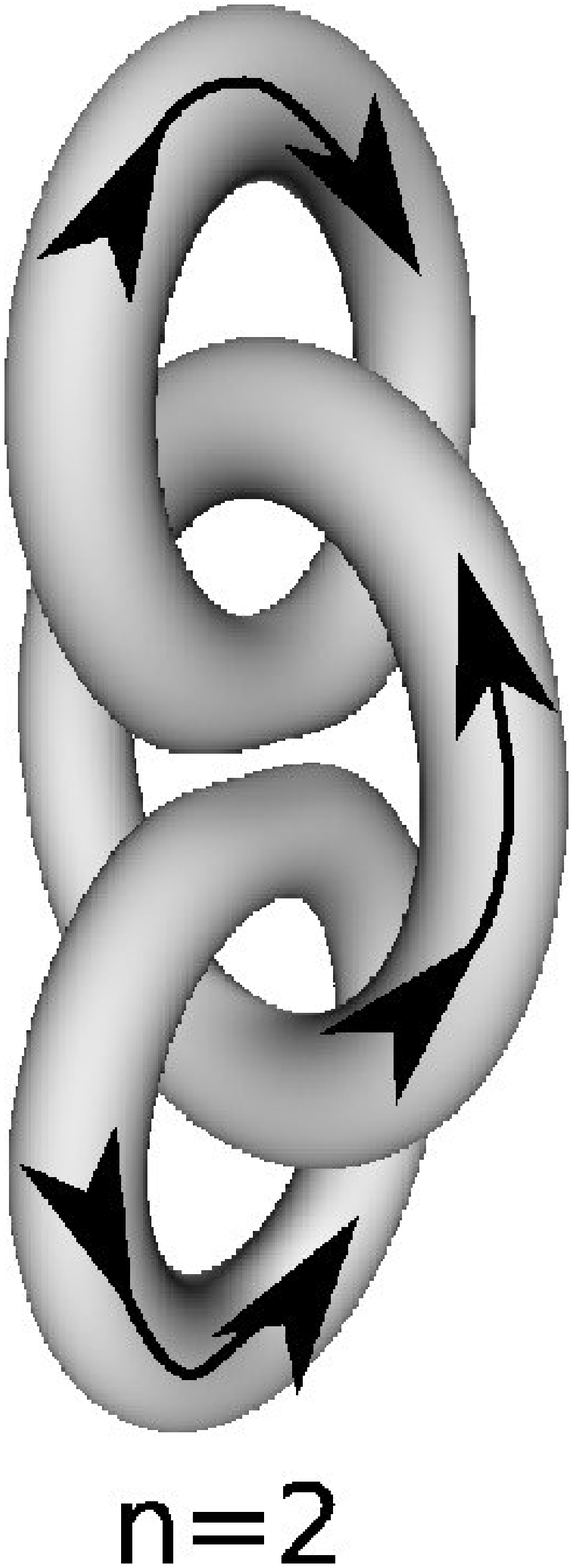} \qquad
\includegraphics[width=.3\columnwidth]{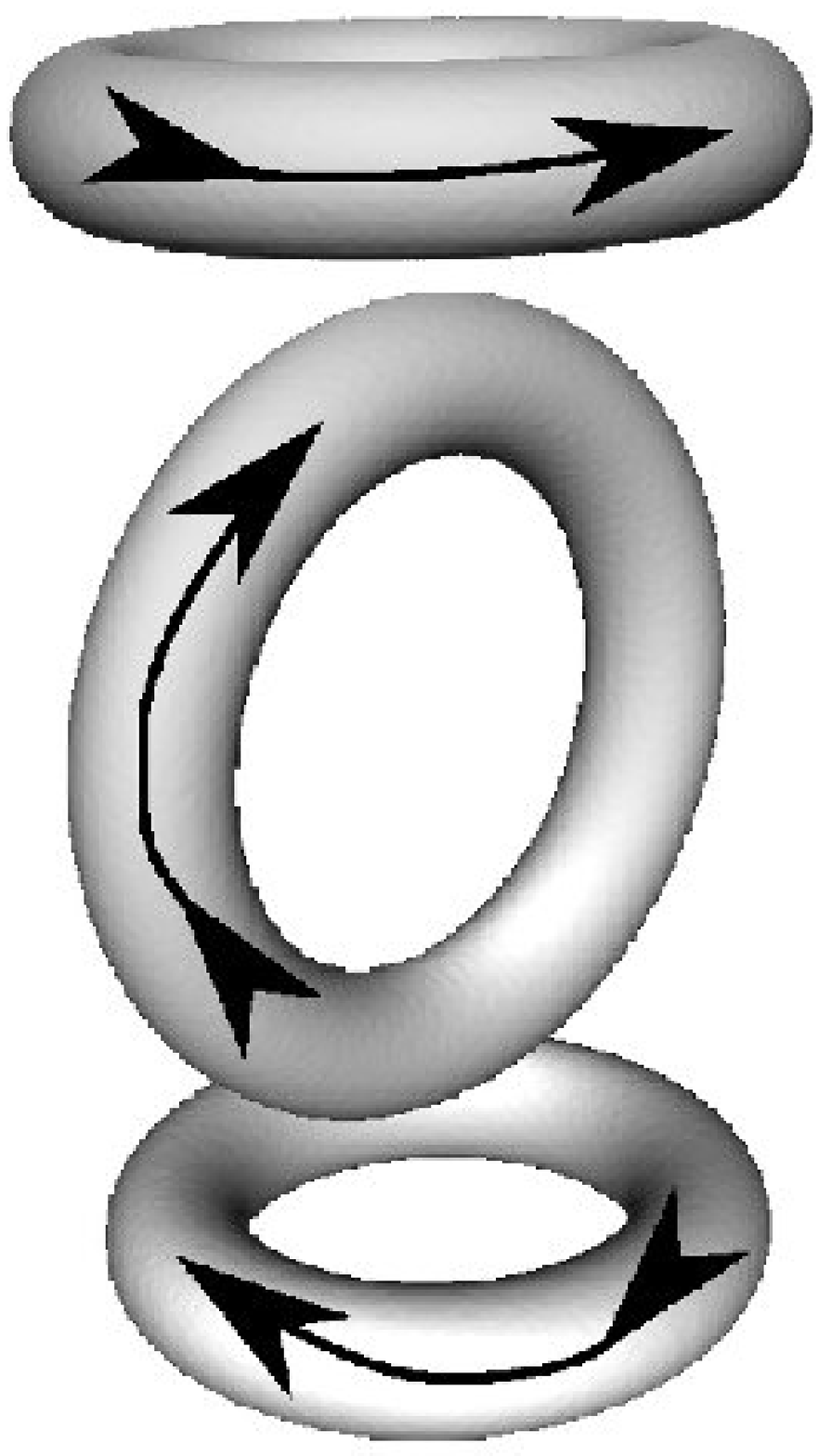}
\end{center}\caption[]{
Visualization of the triple ring configuration at the initial time.
Arrows indicate the direction of the field lines in the rings, corresponding
to a configuration with $n=0$ (left) and $n=2$ (center).
On the right the non-interlocked configuration with $n=0$ is shown.
}\label{flux_configuration}\end{figure}

The situation becomes more interesting when we consider a flux
configuration that is interlocked, but with zero linking number.
This can be realized quite easily by considering a configuration of
two interlocked flux rings where a third flux ring is connected with
one of the other two rings such that the total linking number becomes
either 0 or 2, depending on the relative orientation of the additional ring,
as is illustrated in \Fig{flux_configuration}.
Topologically, the configuration with linking numbers of 0 and 2 are the
same except that the orientation of the field lines in the upper ring
is reversed.
Nevertheless, the simple topological interpretation becomes problematic
in the case of zero linking number,
because then also the magnetic helicity is zero, so the bound of $M$ from below
disappears, and $M$ can now in principle freely decay to zero.
One might expect that the topology should then still be preserved,
and that the linking number as defined above, which is a quadratic
invariant, should be replaced with a higher order invariant
\cite{RA94,HM02,Kom09}.
It is also possible that in a topologically interlocked configuration
with zero linking number the magnetic helicity spectrum $H(k)$ is still
finite and that bound \eq{bound} may still be meaningful.
In order to address these questions we perform numerical simulations
of the resistive magnetohydrodynamic equations using simple interlocked
flux configurations as initial conditions. We also perform a control run
with a non-interlocked configuration and zero helicity in order to compare
the magnetic energy decay with the interlocked case.

Magnetic helicity evolution is independent of the equation of state and
applies hence to both compressible and incompressible cases.
In agreement with earlier work \cite{KB99} we assume an isothermal gas,
where pressure is proportional to density and the sound speed is constant.
However, in all cases the bulk motions stay subsonic, so for all practical
purposes our calculations can be considered nearly incompressible, which
would be an alternative assumption that is commonly made \cite{GM00}.

\section{Model}

We perform simulations of the resistive magnetohydrodynamic equations
for a compressible isothermal gas where the pressure is given by
$p=\rho\cs^2$, with $\rho$ being the density and $\cs$ being
the isothermal sound speed.
We solve the equations for $\AAA$, the velocity $\UU$,
and the logarithmic density $\ln\rho$ in the form
\EQ
{\partial\AAA\over\partial t}=\UU\times\BB+\eta\nabla^2\AAA,
\label{dAdt}
\EN
\EQ
{\DD\UU\over\DD t}=-\cs^2\nab\ln\rho+\JJ\times\BB/\rho+\FF_{\rm visc},
\label{dUdt}
\EN
\EQ
{\DD\ln\rho\over\DD t}=-\nab\cdot\UU,
\label{dlnrhodt}
\EN
where
$\FF_{\rm visc}=\rho^{-1}\nab\cdot2\nu\rho\SSSS$ is the viscous force,
$\SSSS$ is the traceless rate of strain tensor, with components
${\sf S}_{ij}=\half(U_{i,j}+U_{j,i})-\onethird\delta_{ij}\nab\cdot\UU$,
$\JJ=\nab\times\BB/\mu_0$ is the current density,
$\nu$ is the kinematic viscosity, and $\eta$ is the magnetic diffusivity.

The initial magnetic field is given by a suitable arrangement
of magnetic flux ropes, as already illustrated in \Fig{flux_configuration}.
These ropes have a smooth Gaussian cross-sectional profile that can easily be
implemented in terms of the magnetic vector potential.
We use the {\textsc Pencil Code}
(\texttt{http://pencil-code.googlecode.com}),
where this initial condition for $\AAA$ is
already prepared, except that now we adopt a configuration consisting of
three interlocked flux rings (\Fig{flux_configuration}) where the linking number
can be chosen to be either 0 or 2, depending only on the field
orientation in the last (or the first) of the three rings.
Here, the two outer rings have radii $R_{\rm o}$, while the inner ring
is slightly bigger and has the radius $R_{\rm i}=1.2R_{\rm o}$,
but with the same flux.
We use $R_{\rm o}$ as our unit of length.
The sound travel time is given by $T_{\rm s}=R_{\rm o}/\cs$.

In the initial state we have $\UU=\bm{0}$ and $\rho=\rho_0=1$.
Our initial flux, $\Phi=\int\BB\cdot\dd\SSS$, is the same for all tubes
with $\Phi=0.1\,\cs R_{\rm o}^2\sqrt{\mu_0\rho_0}$.
This is small enough for compressibility effects to be unimportant,
so the subsequent time evolution is not strongly affected by this choice.
For this reason, the Alfv\'en time,
$T_{\rm A}=\sqrt{\mu_0\rho_0}R_{\rm o}^3/\Phi$,
will be used as our time unit.
In all our cases we have $T_{\rm A}=10T_{\rm s}$ and
denote the dimensionless time as $\tau=t/T_{\rm A}$.
In all cases we assume that the magnetic Prandtl number, $\nu/\eta$, is unity,
and we choose $\nu=\eta=10^{-4}R_{\rm  o}\cs=10^{-3}R_{\rm  o}^2/T_{\rm A}$.
We use $256^3$ mesh points.

We have chosen a fully compressible code, because it is readily available to us.
Alternatively, as discussed at the end of \Sec{Introduction},
one could have chosen an incompressible code by ignoring the continuity
equation and computing the pressure such that $\nab\cdot\UU=0$ at all times.
Such an operation breaks the locality of the physics and is computationally
more intensive, because it requires global communication.

\section{Results}

\begin{figure}[t!]\begin{center}
\includegraphics[width=.3\columnwidth]{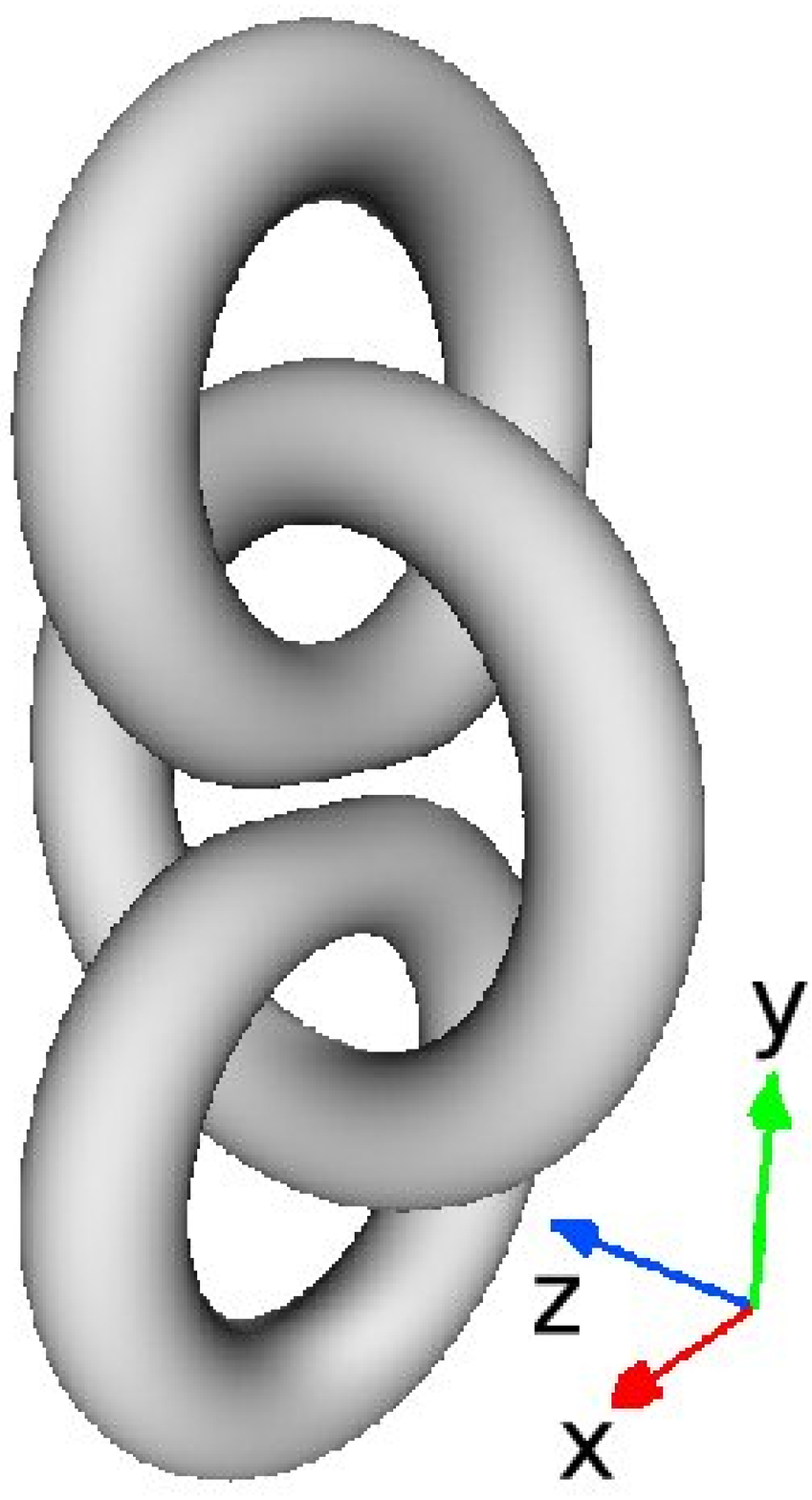}
\includegraphics[width=.3\columnwidth]{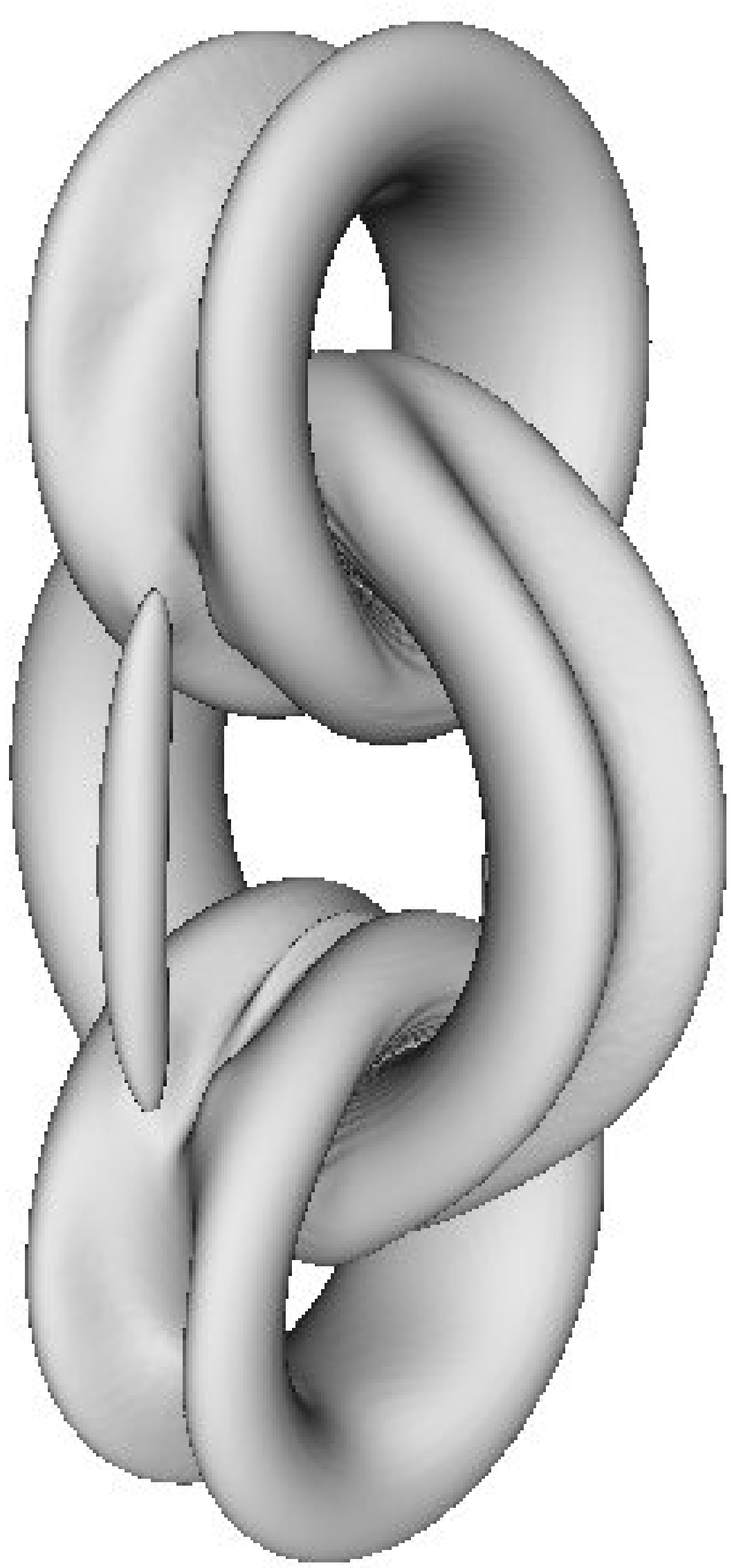}
\includegraphics[width=.3\columnwidth]{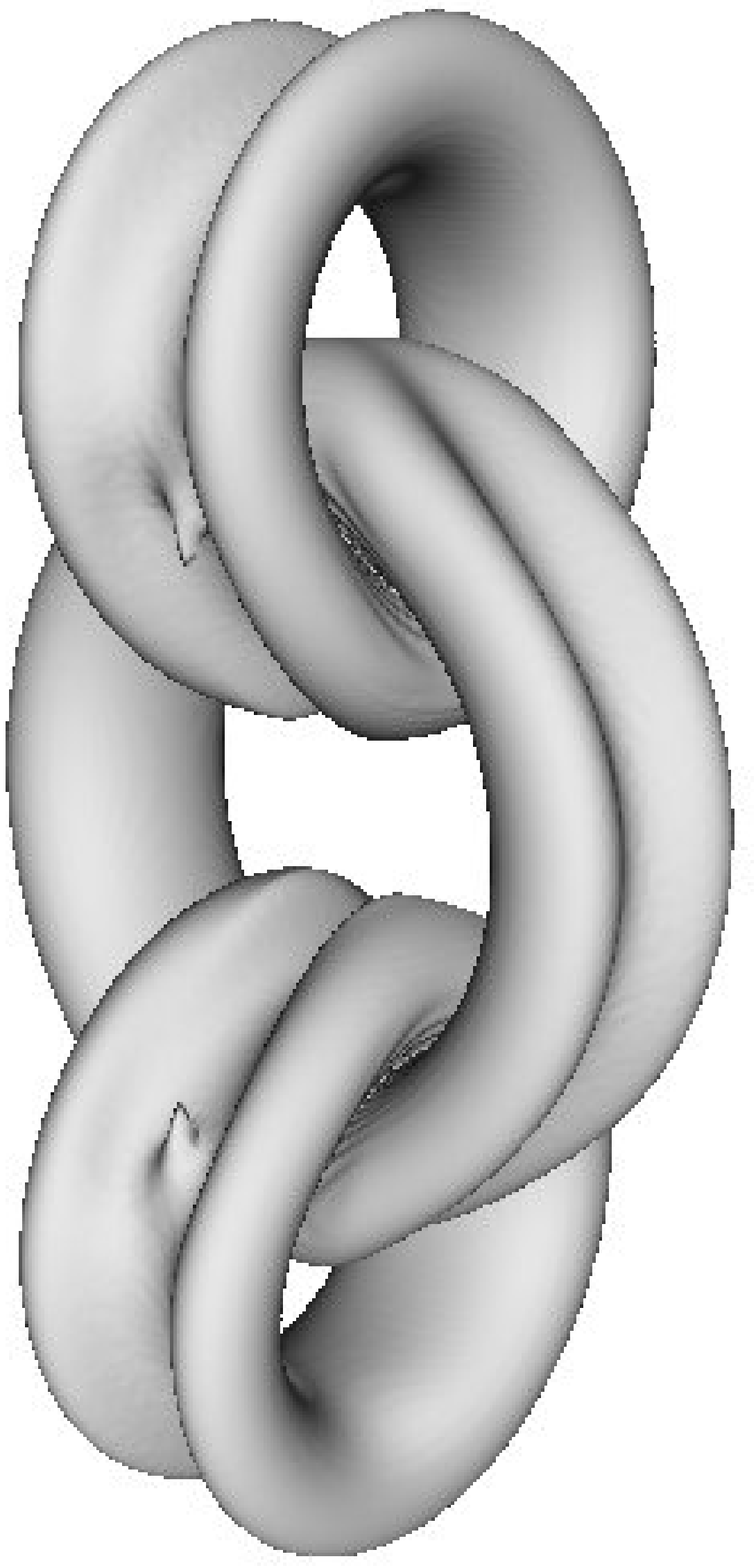}
\end{center}\caption[]{
Visualization of the triple ring configuration at $\tau=0$ (left), as well as
at $\tau=0.5$ with
zero linking number (center) and finite linking number (right).
The three images are in the same scale.
The change in the direction of the field in the upper ring
gives rise to a corresponding change
in the value of the magnetic helicity.
In the center we can see the emergence of a new flux ring encompassing
the two outer rings. Such a ring is not seen on the right.
}\label{256b2_t0_t5}\end{figure}

Let us first discuss the visual appearance of the three interlocked
flux rings at different times.
In \Fig{256b2_t0_t5} we compare the three rings for the zero and finite
magnetic helicity cases at the initial time and at $\tau=0.5$.
Note that each ring shrinks as a result of the tension force.
This effect is strongest in the core of each ring, causing the
rings to show a characteristic indentation that was also seen in
earlier inviscid and non-resistive
simulations of two interlocked flux rings \cite{KB99}.

\begin{figure}[t!]\begin{center}
\includegraphics[width=\columnwidth]{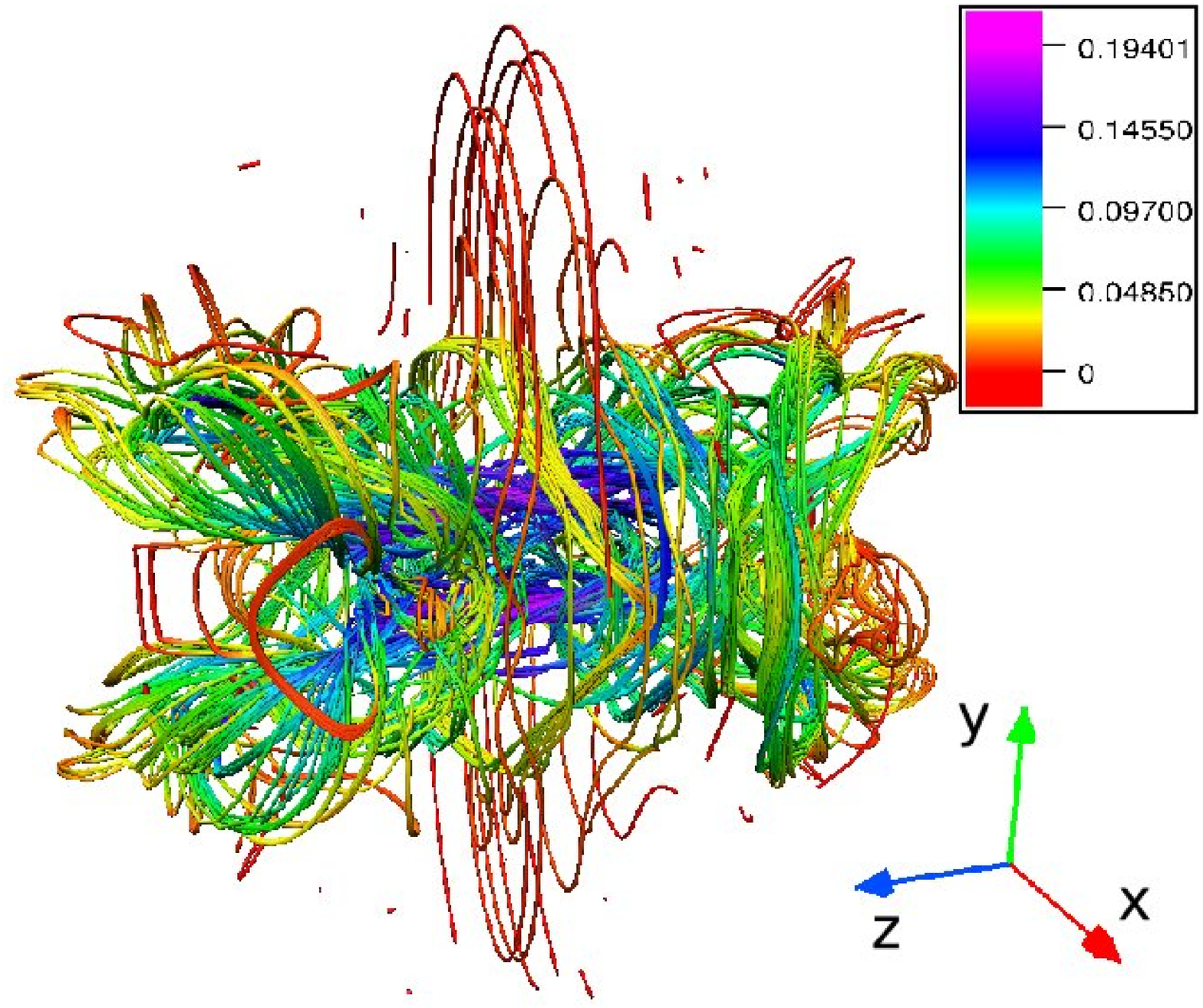} \\
\includegraphics[width=\columnwidth]{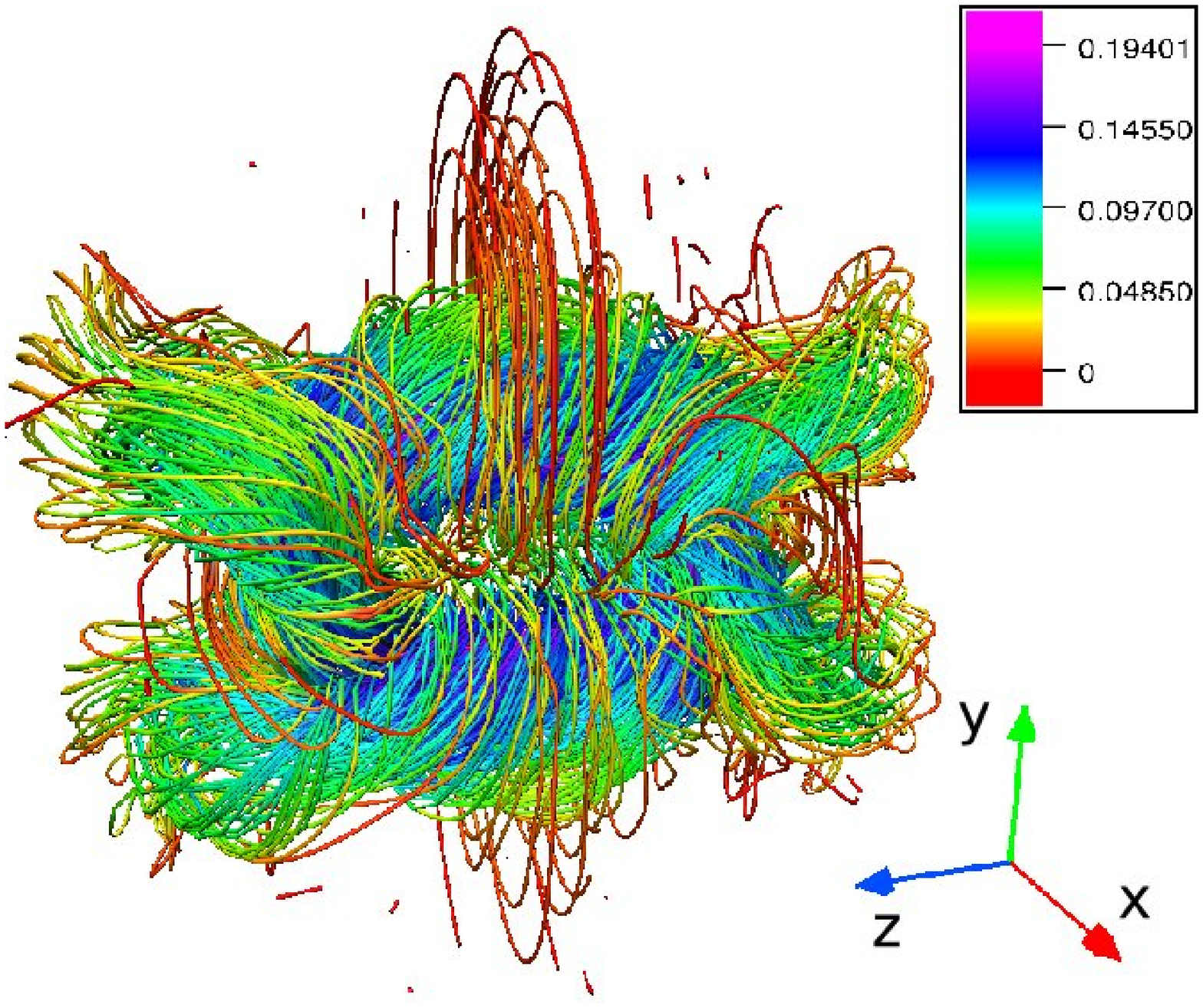}
\end{center}\caption[]{
(Color online) Magnetic flux tubes at time $\tau=4$
for the case of zero linking
number (upper picture) and finite linking number (lower picture). The colors
represent the magnitude of the magnetic field, where the scale goes from red
(lowest) over green to blue (highest).
}\label{field_tubes}\end{figure}

\begin{figure*}[t!]\begin{center}
\includegraphics[width=\textwidth]{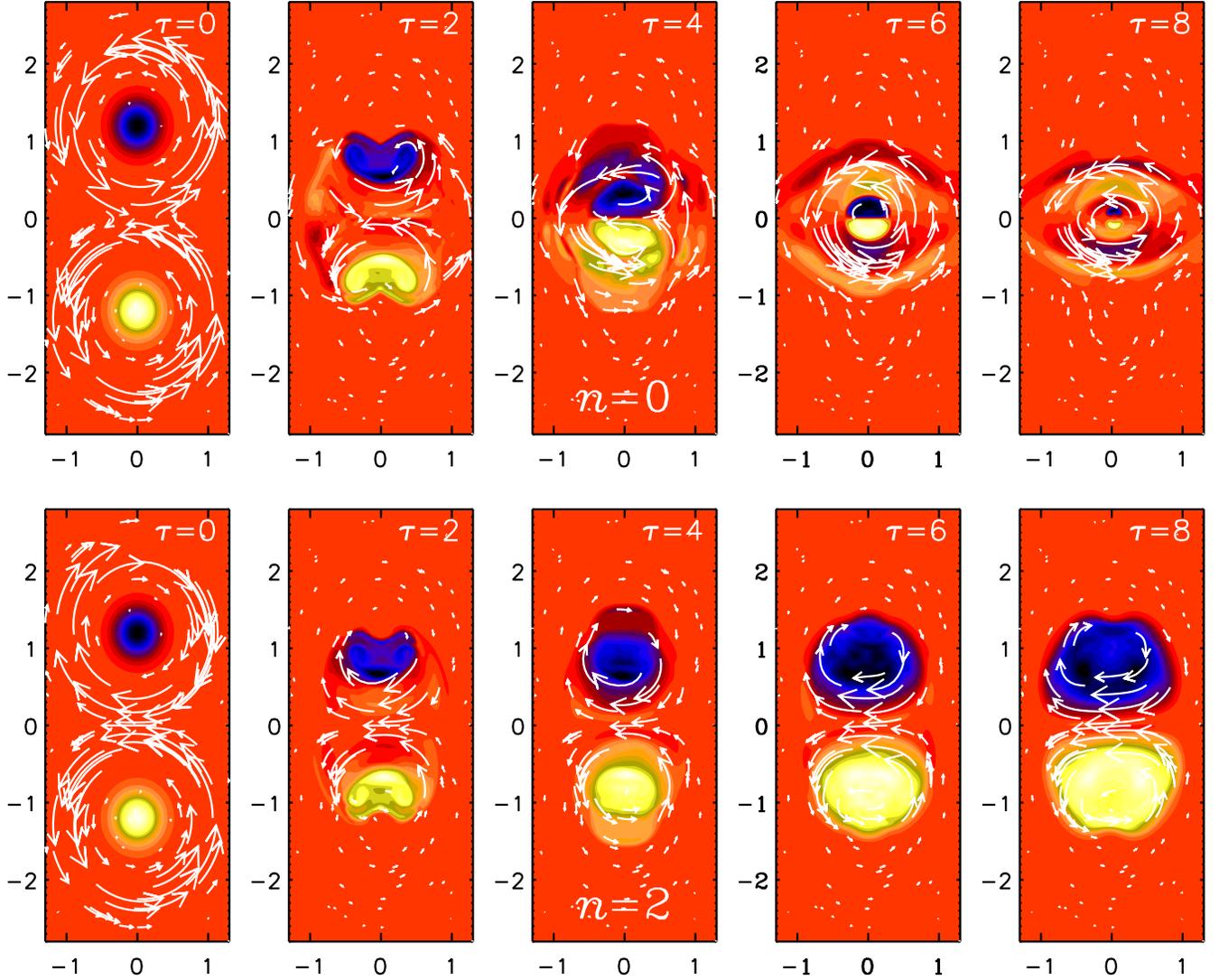}
\end{center}\caption[]{
(Color online) Cross-sections in the $xy$ plane of the magnetic field
with zero linking number (upper row)
and finite linking number (lower row).
The $z$ component (pointing out of the plane)
is shown together with vectors of the field in the plane.
Light (yellow) shades indicate positive values and
dark (blue) shades indicate negative values.
Intermediate (red) shades indicate zero value.
}\label{pbcomp}\end{figure*}

At early times, visualizations of the field show little difference,
but at time $\tau=0.5$ some differences emerge in that the configuration
with zero linking number develops an outer ring encompassing the
two rings that are connected via the inner ring; see \Fig{256b2_t0_t5}.
This outer ring is absent in the configuration with finite linking number.

The change in topology becomes somewhat clearer if we plot the magnetic field
lines (see \Fig{field_tubes}).
For the $n = 2$ configuration, at time $\tau=4$
one can still see a structure of three interlocked rings,
while for the $n = 0$ case no clear structure can be recognized. Note that the
magnitude of the magnetic field has diminished more strongly for $n = 0$ than
for $n = 2$. This is in accordance with our initial expectations.

The differences between the two configurations become harder to interpret
at later times.
Therefore we compare in \Fig{pbcomp} cross-sections of the magnetic
field for the two cases.
The $xy$ cross-sections show clearly the development of
the new outer ring in the zero linking number configuration.
From this figure it is also evident that the zero linking number case
suffers more rapid decay because of the now {\it anti-aligned}
magnetic fields (in the upper panel $B_x$ is of opposite sign about
the plane $y=0$ while it is negative in the lower panel).

\begin{figure}[t!]\begin{center}
\includegraphics[width=\columnwidth]{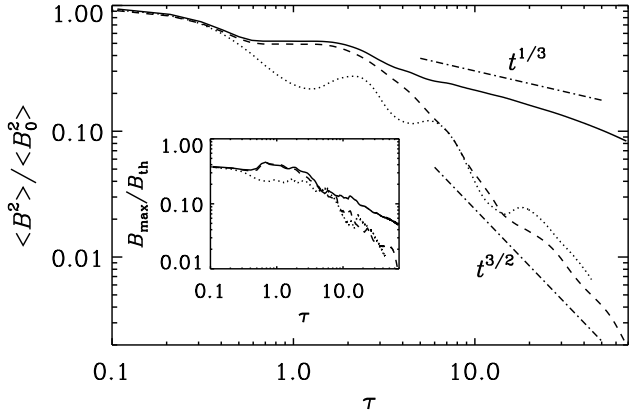}
\end{center}\caption[]{
Decay of magnetic energy (normalized to the initial value)
for linking numbers 2 (solid line) and 0 (dashed line).
The dotted line gives the decay for a control run with
non-interlocked rings.
The dash-dotted lines indicate $t^{1/3}$ and $t^{3/2}$
scalings for comparison.
The inset shows the evolution of the maximum field strength
in units of the thermal equipartition value,
$B_{\rm  th}=\cs(\rho_0\mu_0)^{1/2}$.
}\label{pcomp_mag}\end{figure}

The evolution of magnetic energy is shown in \Fig{pcomp_mag} for
the cases with zero and finite linking numbers.
Even at the time $\tau\approx0.6$, when the rings have just come into mutual
contact, there is no clear difference in the decay for the two cases.
Indeed, until the time $\tau\approx2$ the magnetic energy evolves still
similarly in the two cases, but then there is a pronounced difference where
the energy in the zero linking number case shows a rapid decline
(approximately like $t^{-3/2}$), while in the case with finite linking
number it declines much more slowly (approximately like $t^{-1/3}$).
However, power law behavior is only expected under turbulent conditions
and not for the relatively structured field configurations considered here.
The energy decay in the zero linking number case is roughly the same
as in a case of three flux rings that are not interlocked.
The result of a corresponding control run is shown as a dotted line
in \Fig{pcomp_mag}.
At intermediate times, $0.5<\tau<5$, the magnetic energy of the
control run has diminished somewhat faster than in the interlocked
case with $n=0$.
It is possible that this is connected with the interlocked nature of the
flux rings in one of the cases.
Alternatively, this might reflect the presence of rather different
dynamics in the non-interlocked case, which seems to be strongly controlled
by oscillations on the Alfv\'en time scale.
Nevertheless, at later times the decay laws are roughly the same
for non-interlocked and interlocked non-helical cases.

\begin{figure}[t!]\begin{center}
\includegraphics[width=\columnwidth]{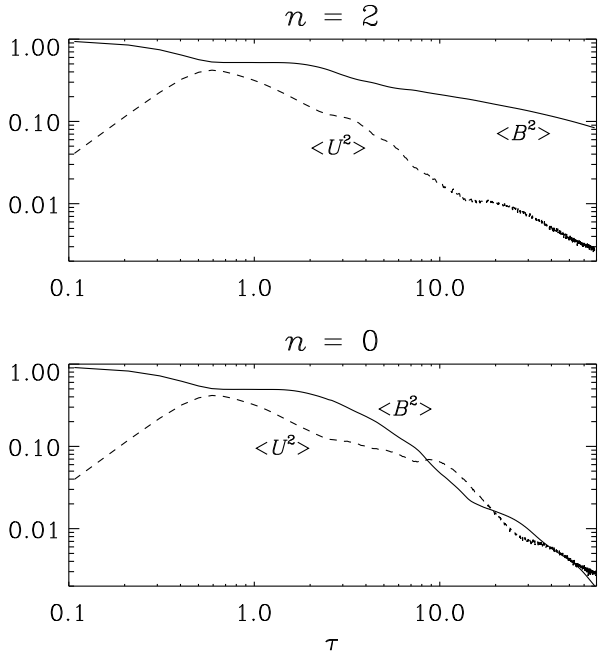}
\end{center}\caption[]{
Comparison of the evolution of kinetic and magnetic energies
in the cases with finite and with vanishing linking numbers.
Note that in both cases the maximum kinetic energy is reached
at the time $\tau\approx0.6$.
The two cases begin to depart from each other after $\tau\approx2$.
In the non-helical case the magnetic energy shows a sharp drop and
reaches equipartition with the kinetic energy, while in the helical case
the magnetic energy stays always above the equipartition value.
}\label{pcomp_kin}\end{figure}

The time when the rings come into mutual contact is marked by a
maximum in the kinetic energy at $\tau\approx0.6$.
This can be seen from \Fig{pcomp_kin}, where we compare kinetic and
magnetic energies separately for the cases with finite and zero
linking numbers.
Note also that in the zero-linking number case magnetic and kinetic
energies are nearly equal and decay in the same fashion.

\begin{figure}[t!]\begin{center}
\includegraphics[width=\columnwidth]{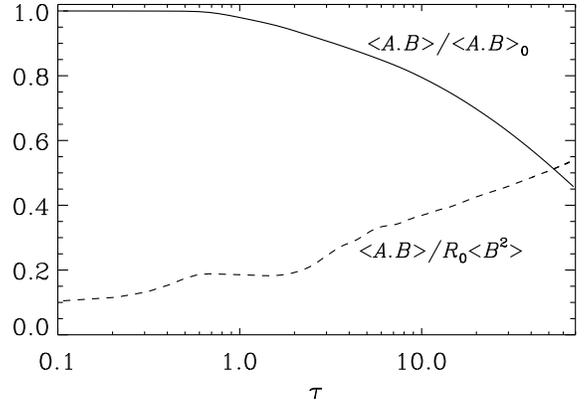}
\end{center}\caption[]{
Evolution of magnetic helicity in the case with finite linking number.
In the upper panel, $\bra{\AAA\cdot\BB}$ is normalized to its initial value
(indicated by subscript 0) while in the lower panel it is normalized to the
magnetic energy divided by $R_{\rm o}$.
}\label{pcomp_hel}\end{figure}

Next we consider the evolution of magnetic helicity in \Fig{pcomp_hel}.
Until the time $\tau\approx0.6$ the value of the magnetic helicity has hardly
changed at all.
After that time there is a gradual decline, but it is slower than
the decline of magnetic energy.
Indeed, the ratio $\bra{\AAA\cdot\BB}/\bra{\BB^2}$, which corresponds to
a length scale, shows a gradual increase from $0.1R_{\rm o}$ to nearly
$0.6R_{\rm o}$ at the end of the simulation.
This reflects the fact that the field has become smoother and more
space-filling with time.

\begin{figure}[t!]\begin{center}
\includegraphics[width=\columnwidth]{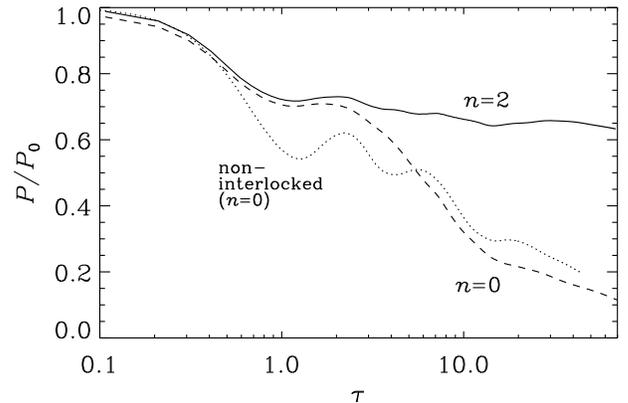}
\end{center}\caption[]{
Decay of the unsigned magnetic flux $P$ (normalized to the initial
value $P_0$) for the cases with $n=0$ and $n=2$.
The dotted line gives the decay for a control run with
non-interlocked rings.
}\label{pcomp_flux}\end{figure}

Given that the magnetic helicity decays only rather slowly,
one must expect that the fluxes $\Phi_i$ of the three rings
also only change very little.
Except for simple configurations where flux tubes
are embedded in field-free regions, it is in general difficult to
measure the actual fluxes, as defined in \Eq{FluxDef}.
On the other hand,
especially in observational solar physics, one often uses the so-called
{\it unsigned} flux \cite{Zwa85,SH94}, which is defined as
\EQ
P_{\rm 2D}=\int_S|\BB|\,d S.
\EN
For a ring of flux $\Phi$ that intersects the surface in the middle
at right angles the net flux cancels to zero, but the unsigned flux
gets contributions from both intersection, so $P_{\rm 2D}=2|\Phi|$.
In three-dimensional simulations it is convenient to determine
\EQ
P=\int_V|\BB|\,d V.
\EN
For several rings, all with radius $R$, we have
\EQ
P=2\pi R \sum_{i=1}^N |\Phi_i|=\pi N R P_{\rm 2D},
\EN
where $N$ is the number of rings.
In \Fig{pcomp_flux} we compare the evolution of $P$ (normalized to the
initial value $P_0$) for the cases with $n=0$ and $n=2$.
It turns out that after $\tau=1$ the value of $p$
is nearly constant for $n=2$, but not for $n=0$.

\begin{figure}[t!]\begin{center}
\includegraphics[width=\columnwidth]{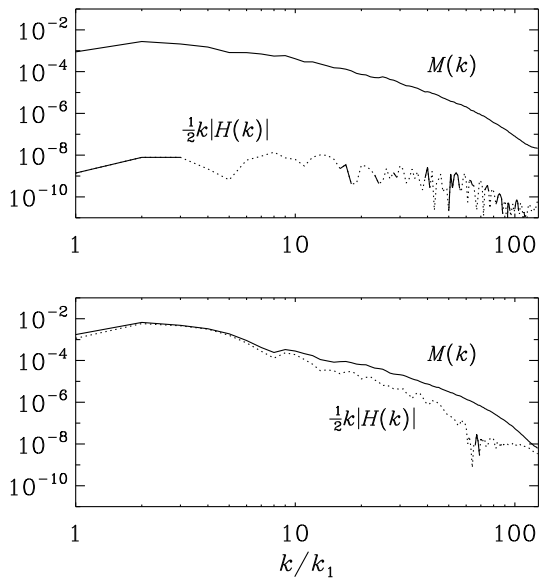}
\end{center}\caption[]{
Comparison of spectra of magnetic energy and magnetic helicity
in the case with zero linking number (upper panel) and finite
linking number (lower panel) at $\tau=5$.
Stretches with negative values of $H(k)$ are shown as dotted lines.
}\label{pspec_comp}\end{figure}

Let us now return to the earlier question of whether
a flux configuration with zero linking number can have finite spectral
magnetic helicity, i.e.\ whether $H(k)$ is finite but of opposite
sign at different values of $k$.
The spectra $M(k)$ and $H(k)$ are shown in \Fig{pspec_comp} for
the two cases at time $\tau=5$.
This figure shows that in the configuration with zero linking number
$H(k)$ is essentially zero for all values of $k$.
This is not the case and, in hindsight, is hardly expected;
see \Fig{pspec_comp} for the spectra of $M(k)$ and $k|H(k)|/2\mu_0$
in the two cases at $\tau=5$.
What might have been expected is a segregation of helicity not in
the wave-number space, but in the physical space for positive and negative
values of $y$.
It is then possible that magnetic helicity has been destroyed by
locally generated magnetic helicity fluxes between the two domains
in $y>0$ and $y<0$.
However, this is not pursued further in this paper.

\begin{figure}[t!]\begin{center}
\includegraphics[width=\columnwidth]{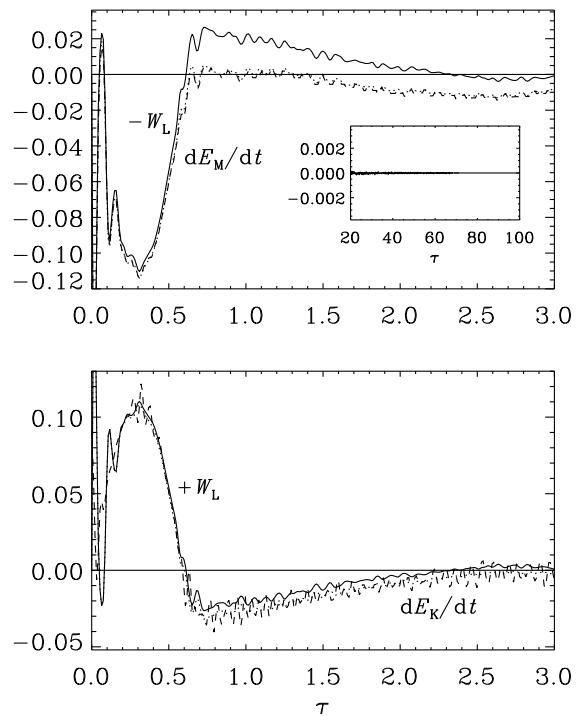}
\end{center}\caption[]{
Evolution of the rate of work done against the Lorentz force, $-W_{\rm L}$,
together with $\dd E_{\rm M}/\dd t$ (upper panel), as well as
the rate of work done by the Lorentz force, $+W_{\rm L}$,
together with $\dd E_{\rm K}/\dd t$ (lower panel),
all normalized in units of $E_{\rm M}/T_{\rm s}$,
for the case with finite linking number.
The inset shows $-W_{\rm L}$ at late times for the case with
$n=0$ (solid line) and $n=2$ (dashed line).
}\label{pdecay_H4_256d2}\end{figure}

In order to understand in more detail the way the energy is dissipated,
we plot in \Fig{pdecay_H4_256d2} the evolution of the time derivative of
the magnetic energy $E_{\rm M}={1\over2\mu_0}\int\BB^2\,d V$ (upper panel)
and the kinetic energy $E_{\rm K}=\half\int\rho\UU^2\,d V$ (lower panel).
In the lower panel we also show the rate of work done by the Lorentz force,
$W_{\rm L}=\int\UU\cdot(\JJ\times\BB)\,d V$, and in the upper panel we
show the rate of work done against the Lorentz force, $-W_{\rm L}$.
All values are normalized by $E_{\rm M0}/T_{\rm s}$, where $E_{\rm M0}$
is the value of $E_{\rm M}$ at $\tau=0$.

The rates of magnetic and kinetic energy dissipation, $\epsilon_{\rm M}$
and $\epsilon_{\rm K}$, respectively, can be read off as the difference
between the two curves in each of the two panels in \Fig{pdecay_H4_256d2}.
Indeed, we have
\EQ
-W_{\rm L}-d E_{\rm M}/d t=\epsilon_{\rm M},
\EN
\EQ
W_{\rm L}+W_{\rm C}-d E_{\rm K}/d t=\epsilon_{\rm K},
\EN
where the compressional work term
$W_{\rm C}=\int p\nab\cdot\UU\,\dd V$ is found to be negligible
in all cases.
Looking at \Fig{pdecay_H4_256d2} we can say that
at early times ($0<\tau<0.7$) the magnetic field contributes to driving
fluid motions ($W_{\rm L}>0$) while at later times some of the
magnetic energy is replenished by kinetic energy ($W_{\rm L}<0$),
but since magnetic energy dissipation still dominates, the
magnetic energy is still decaying ($d E_{\rm M}/d t<0)$.
The maximum dissipation occurs around the time $\tau=0.7$.
The magnetic energy dissipation is then about twice  as large as the
kinetic energy dissipation.
We note that the ratio between magnetic and kinetic energy
dissipations should also depend on the value of the magnetic Prandtl number,
$\Pm=\nu/\eta$, which we have chosen here to be unity.
In this connection it may be interesting to recall that one finds similar
ratios of $\epsilon_{\rm K}$ and $\epsilon_{\rm M}$
both for helical and non-helical turbulence \cite{HBD03}.
At smaller values of $\Pm$ the ratio of $\epsilon_{\rm K}$ to
$\epsilon_{\rm K}+\epsilon_{\rm M}$ diminishes like $\Pm^{-1/2}$
for helical turbulence \cite{B09}.
In the present case the difference between $n=0$ and 2 is, again, small.
Only at later times there is a small difference in $W_{\rm L}$, as is
shown in the inset of \Fig{pdecay_H4_256d2}.
It turns out that, for $n=2$, $W_{\rm L}$ is positive while for $n=0$
its value fluctuates around zero.
This suggests that the $n=2$ configuration is able to sustain fluid motions
for longer times than the $n=0$ configuration.
This is perhaps somewhat unexpected, because the helical configuration
($n=2$) should be more nearly force free than the non-helical configuration.
However, this apparent puzzle is simply explained by the fact that the
$n=2$ configuration has not yet decayed as much as the $n=0$ configuration has.

\section{Conclusions}

The present work has shown that the rate of magnetic energy dissipation
is strongly constrained by the presence of magnetic helicity and not
by the qualitative degree of knottedness.
In our example of three interlocked flux rings we considered two flux chains,
where the topology is the same
except that the relative orientation of the magnetic field is reversed
in one case.
This means that the linking number switches from 2 to 0, just depending
on the sign of the field in one of the rings.
The resulting decay rates are dramatically different in the two cases,
and the decay is strongly constrained in the case with finite magnetic
helicity.

The present investigations reinforce the importance of considering
magnetic helicity in studies of reconnection.
Reconnection is a subject that was originally considered in
two-dimensional studies of X-point reconnection \cite{Par57,Bis86}.
Three-dimensional reconnection was mainly considered in the last 20 years.
An important aspect is the production of current sheets in the
course of field line braiding \cite{Ber93}.
Such current sheets are an important contributor to coronal heating
\cite{GN96}.
The crucial role of magnetic helicity has also been recognized in several
papers \cite{Hu97,Liu07}.
However, it remained unclear whether the decay of interlocked flux
configurations with zero helicity might be affected by the degree of
tangledness.
Our present work suggests that a significant amount of dissipation should
only be expected from tangled magnetic fields that have zero or small
magnetic helicity, while tangled regions with finite magnetic helicity
should survive longer and are expected to dissipate less efficiently.

\acknowledgments
We acknowledge the allocation of computing resources provided by the
Swedish National Allocations Committee at the Center for
Parallel Computers at the Royal Institute of Technology in
Stockholm and the National Supercomputer Centers in Link\"oping.
This work was supported in part by
the European Research Council under the AstroDyn Research Project No.\ 227952
and the Swedish Research Council Grant No.\ 621-2007-4064.

\end{document}